\documentclass[12pt]{iopart}

\usepackage{graphicx}
\usepackage[hypertex]{hyperref}
\begin{document}

\title[Thermodynamics of elementary excitations in artificial magnetic square ice]{Thermodynamics of elementary excitations in artificial magnetic square ice }

\author{R.C. Silva$^1$, F.S. Nascimento$^1$, L.A.S. M\'{o}l$^1$, W.A. Moura-Melo$^1$ and A.R. Pereira$^1$}

\ead{apereira@ufv.br} \address{$^1$Departamento de F\'isica,
Universidade Federal de
Vi\c cosa, Vi\c cosa, 36570-000, Minas Gerais, Brazil
}

\submitto{NJP}

\begin{abstract}
We investigate the thermodynamics of artificial square spin ice systems assuming only dipolar interactions among the islands that compose the array. The emphasis is given on the effects of the temperature on the elementary excitations (magnetic monopoles and their strings). By using Monte Carlo techniques we calculate the specific heat, the density of poles and their average separation as functions of temperature. The specific heat and average separation between monopoles with opposite charges exhibit a sharp peak and a local maximum, respectively, at the same temperature, $T_{p}\approx 7.2D/k_{B}$ (here, $D$ is the strength of the dipolar interaction and $k_{B}$ is the Boltzmann constant). As the lattice size is increased, the amplitude of these features also increases but very slowly. Really, the specific heat and the maximum in the average separation $d_{max}$ between oppositely charged monopoles increase logarithmically with the system size, indicating that completely isolated charges could be found only at the thermodynamic limit. In general, the results obtained here suggest that, for temperatures $T \geq T_{p}$, these systems may exhibit a phase with separated monopoles, although the quantity $d_{max}$ should not be larger than a few lattice spacings for viable artificial materials.

\end{abstract}

\maketitle

\section{Introduction}

\indent New methods for exploring geometric frustrations in magnetic systems have been developed recently.
Such methods consist in creating arrays of nanomagnets designed to resemble the disordered magnetic state
known as spin ice. They are essentially composed of lithographically defined two-dimensional ($2d$) ferromagnetic nanostructures (elongated permalloy nanoparticles) with single-domain elements organized in diverse types of geometries (square lattice \cite{Wang06}, hexagonal, brickwork \cite{Li10}, kagome \cite{Ladak10,Mengotti10} etc). Since their geometries are determined lithographically, lattice symmetry and topology can be directly controlled, allowing experimental investigation of a vast set of important theoretical models of statistical physics \cite{Baxter}. These artificial magnetic compounds have the potential of increasing our understanding
of disordered matter and may also lead to new technologies. Therefore, artificial spin ices are object of intense theoretical and experimental investigations \cite{Wang06,Li10,Ladak10,Mengotti10,Moller06,Ke08,Mol09,Mol10,Morgan10,Moller09,Zabel09,Budrikis10,Libal11}.

The trouble is that, in artificial spin ice patterns, the magnetization is unaffected by thermal fluctuations because the magnetic islands contain a large number of spins. Despite the fact that the moment configuration is athermal, these artificial materials can be described through an effective thermodynamics formalism \cite{Nisoli07,Nisoli10}; in addition, some works have introduced a predictive notion of effective temperature \cite{Ke08,Nisoli10}. For instance, an external drive, in the form of an agitating magnetic field behaves as a thermal bath and controls the temperature \cite{Ke08,Nisoli10}. Alternatively, this problem was addressed very recently by using a material with an ordering temperature near room temperature \cite{Kapaklis11}; such experimental work on a square lattice in an external magnetic field confirms  a dynamical ''pre-melting" of the artificial spin ice structure at a temperature well below the intrinsic ordering temperature of the island material, creating a spin ice array that has real thermal dynamics of the artificial spins over an extended temperature range \cite{Kapaklis11}. These findings and also other future possibilities make evident that a more detailed analysis of the effects of thermal fluctuations on a lower dimensional spin ice material should be of large interest for a better understanding of these frustrated systems. In particular, it would also be important to know the roles of elementary excitations in the thermodynamic properties of artificial magnetic ices.

The main aim of this work is exactly this investigation. We are interested in the temperature effects on the excitations (``magnetic monopole defects" and their strings). Actually, since the prediction of monopoles in the usual three-dimensional ($3d$) spin ice materials \cite{Castelnovo08} and their experimental detection
\cite{Fennell09,Morris09,Bramwell09,Kadowaki09,Jaubert09}, the search for these objects in artificial compounds has become an important issue \cite{Mol09,Moller09,Morgan10,Ladak10}. The possible existence of these excitations in artificial and controllable systems is of great interest because they could be studied at room temperature and, more important, they could be directly observed with modern experimental techniques. Curiously, in the case of artificial systems, while the square lattice was the first to be produced \cite{Wang06}, the direct observation of magnetic monopole defects and their motion was firstly accomplished in a kagome geometry \cite{Ladak10}. Still, in this kagome lattice, a direct, real-space observation of the interplay of Dirac strings and monopoles was reported by Mengotti \textit{et.al} \cite{Mengotti10}. For a square lattice, the direct observation of such excitations came only afterward because there was a primary experimental problem: until last year, none of such systems had achieved its ground state through thermodynamic equilibrium \cite{Budrikis10}. Despite predictions\cite{Moller06,Mol09,Mol10}, the studies till recently do not have shown a long-range ordered configuration, perhaps because the researchers have used only non-thermal methods to randomize the array. This problem was experimentally solved by Morgan \textit{et. al.} \cite{Morgan10}. These authors have reported that by allowing the magnetic islands to interact as they are gradually formed at room temperature, the artificial square spin ice can be effectively thermalized, allowing it to find its predicted ground state very closely; thus, they could also identify the small departures from the ground state as elementary excitations of the system, at frequencies that follow a Boltzmann law. Subsequently, Magnetic Force Microscopy (MFM) images of a large number of isolated excitations with their string shapes and corresponding moment flip maps were described in square lattices \cite{Morgan10}. Therefore, the experimental results considering magnetic artificial square ices obtained in Ref.\cite{Morgan10} (which demonstrates the thermal ground-state ordering and the elementary excitations) and Ref.\cite{Kapaklis11} (which achieves a thermodynamic melting transition by using a material with ordering temperature near room temperature) lead us to think that more progress on the development of such arrays may become available in the near future, establishing opportunities to experimentally elucidate their real thermodynamics.


\begin{figure}
\includegraphics[angle=0.0,width=7cm]{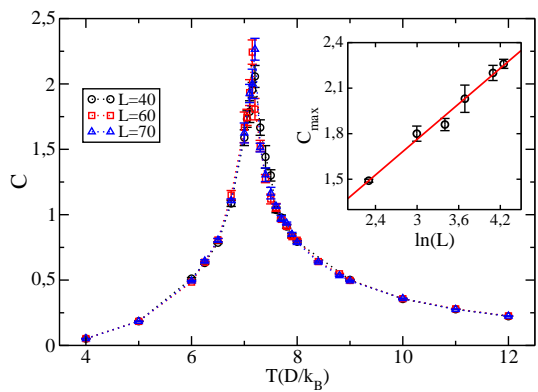}
\caption{ \label{Heat} (Color online) Specific heat as a function of temperature. It exhibits a sharp peak, at a temperature $T_{p}\sim 7.2 D/k_{B}$, which the amplitude increases very slowly with the system size $L$. Inset: the specific heat peak diverges logarithmically with the system size $L$.}
\end{figure}

\section{The model and outlook}

Here, we consider an arrangement of dipoles similar to that experimentally investigated
in Ref. \cite{Wang06}. In our approach, however, the magnetic moment (``spin") of the
island is replaced by an Ising-like point dipole at its center.
In this approach, the internal degrees of freedom of each island are not being considered,
as well as higher order interactions. We expect that this simplification does not 
change significantly the main physical properties of the system. As shown in Ref.~\cite{Leon08},
if the lattice spacing is about two times larger than the island's longest axis, the effect
of higher order interactions is negligible. For smaller lattice spacings the effect of 
higher order interactions is to give more stability for the lowest energy states. In this way
one may expect that as the island size increases, approaching the lattice spacing, the ground-state
should be more robust and the appearance of excitations would cost more energy. While the consideration
of the internal degrees of freedom would reduce the energy scale, the consideration of higher order
interactions would increase it, but none of them are expected to change the physical picture
discussed here.
Thus, in our approach, at each site $(x_{i},y_{i})$ of the square lattice two spin variables are defined: $\vec{S}_{x(i)}$ with components $S_{x}=\pm 1$, $S_{y}=0, S_{z}=0$ located at $\vec{r}_{x}=(x_{i}+1/2,y_{i})$, and $\vec{S}_{y(i)}$ with components $S_{x}=0$, $S_{y}=\pm 1, S_{z}=0$ at $\vec{r}_{y}=(x_{i},y_{i}+1/2)$. Therefore, in a lattice of volume $L^{2}=l^{2}a^{2}$ ($a$ is the lattice spacing) one gets $2\times l^{2}$ spins. Representing the spins of the islands by $\vec{S}_{i}$, which can assume either $\vec{S}_{x(i)}$ or $\vec{S}_{y(i)}$, then the artificial spin ice is described by the
following Hamiltonian
\begin{eqnarray}\label{HamiltonianSI}
H_{SI} &=& Da^{3} \sum_{i\neq j}\left[\frac{\vec{S}_{i}\cdot
\vec{S}_{j}}{r_{ij}^{3}} - \frac{3 (\vec{S}_{i}\cdot
\vec{r}_{ij})(\vec{S}_{j}\cdot \vec{r}_{ij})}{r_{ij}^{5}}\right],
\end{eqnarray}\\
where $D=\mu_{0}\mu^{2}/4\pi a^{3}$ is the coupling constant of
the dipolar interaction. We perform standard Monte Carlo techniques to obtain thermodynamic averages of the system defined by Hamiltonian (\ref{HamiltonianSI}). Periodic boundary conditions were implemented by means of the Ewald Summation \cite{Wang01,Weis03}, used here to avoid spurious results brought about by the use of a cut-off radius\cite{Mol}. Our Monte Carlo procedure comprises a combination of single spin flips and loop moves \cite{Barkema98}, where all spins contained in a closed random loop are flipped according to the Metropolis prescription. In our scheme one Monte Carlo step (MCS) consists of $2\times l^2$ single spin flips and $0.7\times l^2$ worm moves. Usually, $10^4$ MCS were shown to be sufficient to reach equilibrium configurations and we have used $10^5$ configurations to get thermodynamic averages.

\begin{figure}
\includegraphics[angle=0.0,width=8cm]{densidade2.eps}
\caption{ \label{Density} (Color online) Density of pairs of unit-charged monopoles as a function of
temperature. Inset: density of doubly charged monopole pairs. }
\end{figure}

\begin{figure}
\includegraphics[angle=0.0,width=8cm]{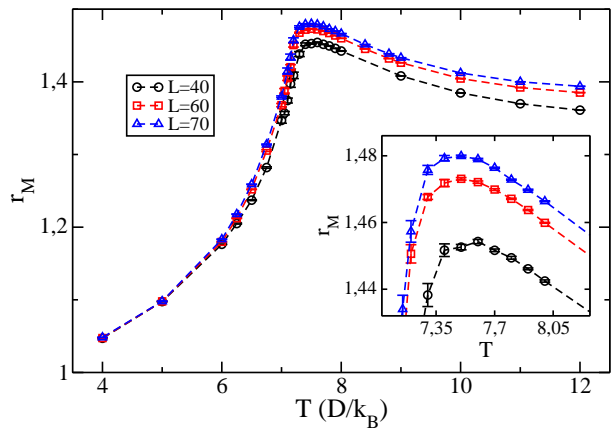}
\caption{ \label{Separation} (Color online) The average separation between charges exhibits a maximum around
the same temperature $T_{p}$ in which the specific heat has a sharp feature. The inset shows, in more details, the region around the maximum.}
\end{figure}

Before presenting the Monte Carlo calculations, it would be interesting to remark on some previous results \cite{Mol09,Moller09,Mol10} and some expectations for these arrays. The ground state configuration of the system in a square lattice is twofold degenerate. If one considers the vorticity in each plaquette, assigning a variable $\sigma=+1$ and $-1$ to clockwise and anticlockwise vorticities respectively, the ground state looks like a checkerboard, with an antiferromagnetic arrangement of the $\sigma$ variable \cite{Mol09,Morgan10}. Of course, the ground state clearly obeys the ice rule (two spins point inward and two point outward in each vertex), but with configurations of topology $1$ (in $2d$, there are two topologies that obey the ice rule. However, they are not degenerate and topology $2$ is more energetic than topology $1$; see Refs.\cite{Wang06,Mol09} for more details). The most elementary excitation is related to the inversion of a single spin (dipole) to generate a localized pair of defects. This is the $3-in$, $1-out$ state in a particular vertex and the $3-out$, $1-in$ state in its adjacent vertex. In principle, these defects could be separated without further violation of the ice rule. Indeed, in our previous papers \cite{Mol09,Mol10}, we have numerically shown that these defects behave as a monopoles pair since their interaction follows a $d=3$ Coulomb law $q/R$, where $q$ measures the strength of the interaction and $R$ is the distance between the poles. However, we have also pointed out that an isolated monopole should be hard to see as effective low-energy degrees of freedom in the $2d$ square spin ice because the background antiferromagnetic order in the ground state confines them \cite{Mol09}, since the ice rule is not degenerate in two dimensions. Actually, in $2d$, there are additional excitations not present in the usual $3d$ spin ice \cite{Castelnovo08}, namely, energetic one-dimensional strings of dipoles (resultant spins at each vertex along a line of adjacent vertices) that terminate in monopoles with opposite charges. Such string excitations could be seen as lines which pass by adjacent vertices that obey the ice rule but sustaining topology $2$ (instead of topology $1$) and hence they cost an energy equal to $b$ times their length $X$, where $b$ is the string tension. When the temperature $T$ of the system is near absolute zero, the shortest path length connecting the monopoles gives the potential energy. The most general expression for the total cost of a pair of monopoles separated by a distance $R$ is the sum of the usual Coulombic term roughly equal to $q/R$, and a term roughly equal to $bX$ resulting from the string joining the monopoles (there is, of course, also a constant term associated with the creation energy of a pair). Note that there is not a unique identification of a given path connecting the ends (monopoles) of the excitation. It is explicitly considered in the fact that the energy is proportional to $X$, which can assume different values for a given $R$. For a sufficiently long string, the string energy is completely dominant; for a short string the Coulomb interaction may have some importance if the size of the end-point monopoles is even smaller (as always occur for these systems). With the above features, these excitations are, to some extent, more similar to Nambu monopoles \cite{Nambu74} than Dirac monopoles. Really, as Nambu suggested, for a modified Dirac monopole theory, the string connecting monopoles has energy and is oriented, having a sense of polarization\cite{Nambu74}.

In the artificial square ices, the ordering causes an anisotropy in the system making the monopoles interaction highly dependent on the direction in which the monopoles are separated in the crystal plane \cite{Mol10}. This anisotropy is manifested in both the Coulomb and linear terms of the potential in such a way that we explicitly write \cite{Mol10}
\begin{eqnarray}\label{Potential}
V(R)=q(\phi)/R+b(\phi)X + c
\end{eqnarray}\\
where $\phi$ is the angle that the line joining the monopole defects makes with the $x$-axis of the array. Numerically, for instance, $q(0) \approx -3.88Da$, $b (0) \approx 9.8 D/a$ while $q(\pi/3) \approx -4.1Da$, $b (\pi/3) \approx 10.1 D/a$. The constant $c\approx 23D$, associated with the pair creation energy \cite{Mol10} ($E_{c}\approx 29D$) is independent of $\phi$. Similar results can be found in the experimental work for the square lattice. Indeed, in Ref.\cite{Morgan10}, the authors have found that, at a temperature $T$, these excitations arise in the system according to the Boltzmann law $\sim \exp(-\beta V(R))$ with $b \approx 10D/a$, $V(a)=E_{c} \approx 30D$ and $\beta=1/k_{B}T$, where $k_{B}$ is the Boltzmann constant. They have also classified the elementary excitations by the number of flipped spins (given by $n$) and a mnemonic character for shape. The three most observed defects are represented by $1$ (a single pair with charges separated by only one lattice spacing) followed by $2L$ (a pair with $n=2$ with the shape of $L$) and $4O$ (an isolated string loop with no charges and having $n=4$ flipped spins) \cite{Morgan10}. Curiously, the second excited state should be $4O$ since its energy is smaller than the energy of $2L$ defect.

\begin{figure}
\includegraphics[angle=0.0,width=7cm]{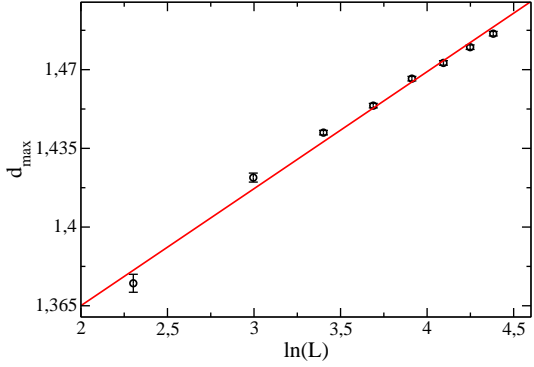}
\caption{ \label{Maximum} (Color online) The maximum of the average separation $d_{max}$ between opposite charges increases logarithmically with the system size $L$.}
\end{figure}


\begin{figure}
\includegraphics[angle=0.0,width=9cm]{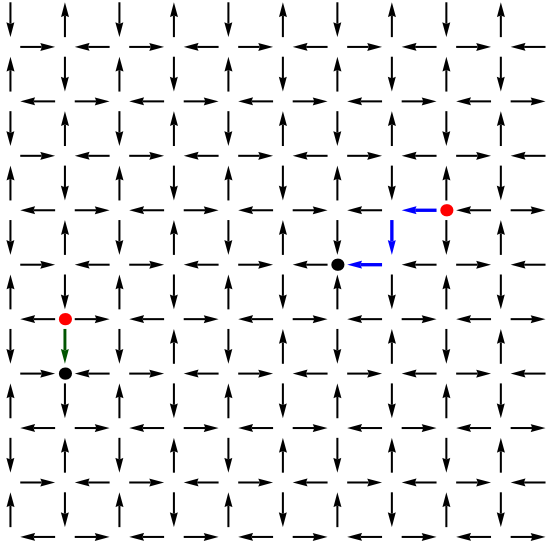}
\caption{ \label{Temp1} (Color online) Snapshot of a particular configuration of excitations for a temperature $T=6.0D/k_{B}$ in a lattice with $L=10a$. Red and black circles are positive and negative charges respectively. In general, for all temperatures below $T_{p}$, each monopole is clearly confined to its counterpart by a string (see the blue arrow indicating the direction of the string for the larger pair. Small pairs (i.e., monopoles bound together tightly in pairs) are indicated by a green arrow.}
\end{figure}


\begin{figure}
\includegraphics[angle=0.0,width=9cm]{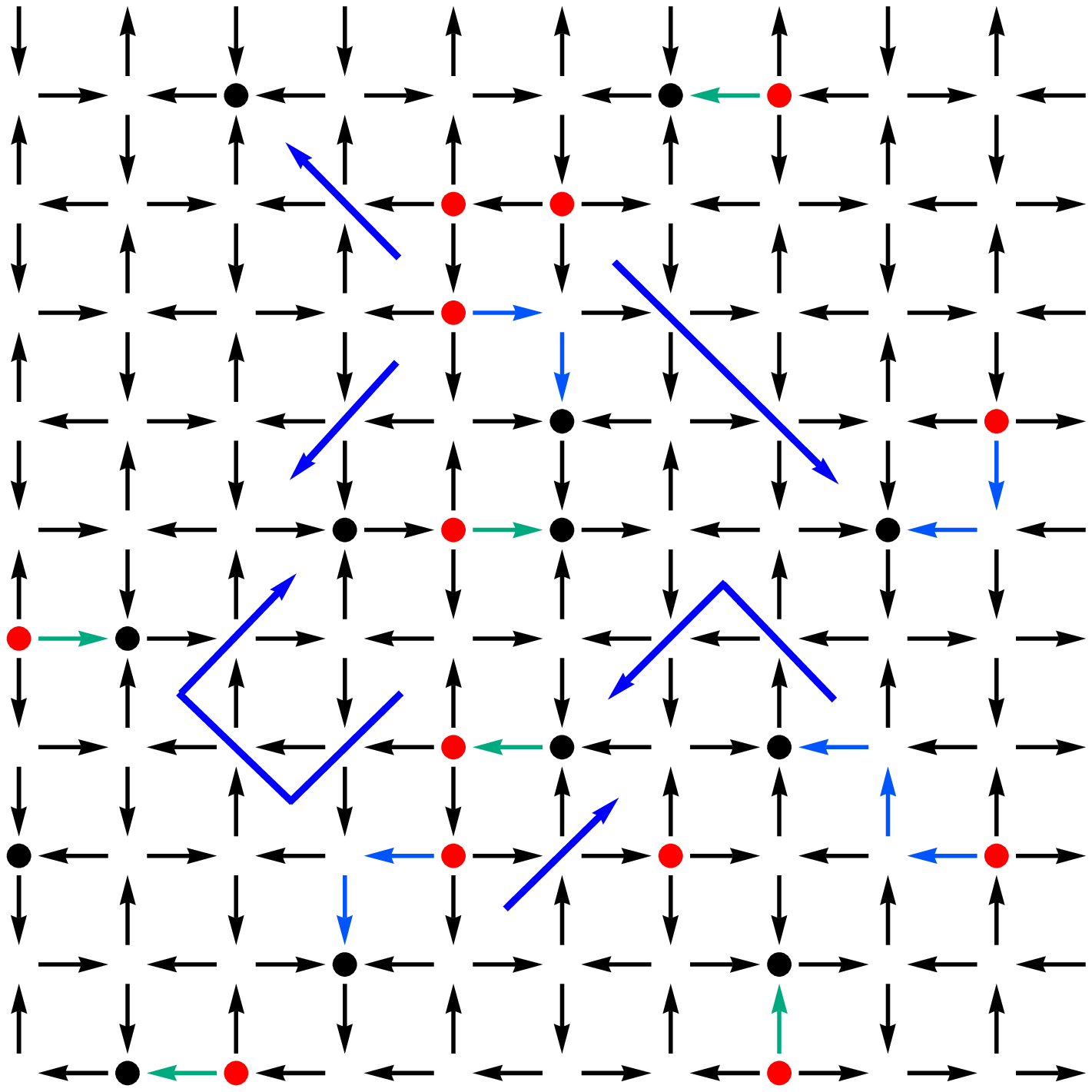}
\caption{ \label{Temp2} (Color online) Snapshot of a particular configuration of excitations for a temperature $T=7.6D/k_{B}$ in a lattice with $L=10a$. Red and black circles are positive and negative charges respectively. For a temperature above $T_{p}$, a small amount of monopoles does not have a string connecting them to their counterparts and, therefore, they seem to be isolated. There are also some pieces of strings (i.e., one-dimensional regions obeying topology $2$, as indicated by blue paths) that do not connect monopoles. Small pairs are indicated by a green arrow.}
\end{figure}

In principle, for the thermodynamics of these systems, the following argument should be
valid: at low temperatures, there is insufficient thermal energy to create long strings (with length $X$ larger than one lattice spacing) and so, the monopoles (with opposite charges) are bound together tightly in pairs. On the other hand, as the temperature is increased, the average separation between the constituents of a pair should also increase, which means that larger strings may become present in the system. Of course, there are several ways of connecting two monopoles by a string of length $X$. Therefore, considering states with $X>>R$, we remember then that the number of configurations for the $m$-step self-avoiding random walk is $N=\delta^{m}$, where $\delta$ is a constant and equal to $3$ for a $2d$ square lattice. For the string with sufficient large $X$, $N$ is well approximated by the random walk result and one obtains $N\simeq \delta^{X/a}$. So the entropy of strings is proportional to $X$, i.e., the many possible ways of connecting two monopoles with a string give rise to a string configurational entropy proportional to $X$. Crudely speaking, then, the string free energy $F=[b-(\ln 3) k_{B}T/a] X$ will imply in an effective string tension $[b-(\ln 3) k_{B}T/a]$ which is positive in the low temperature region and the monopoles are completely confined. Above a certain temperature, it becomes negative, namely, the string looses its tension. The tension decreases like $[b-(\ln 3)k_{B}T/a]$ with increasing $T$, vanishing at some critical temperature $k_{B}T_{c}\approx ba/\ln 3$. Using the average value for the string tension in Eq.(\ref{Potential}), i.e., $b\approx 10D/a$, we then estimate $k_{B}T_{c}\approx 9.1D$. Of course, these theoretical arguments always overestimate the critical temperature. Although this picture leads to a rich physics for this system, predicting free magnetic monopoles and a phase transition, things may be a little more complicated. Really, additionally to the entropic effect discussed just above, there is another entropic contribution which manifests against monopole separation; the monopoles should become close together because it would provide more ways to arrange the surrounding dipoles in the lattice. Such effect introduces a $2d$ Coulombic interaction between the poles, which is proportional to $T$ (i.e., $V_{s}=T\ln(R/a)$). If the temperature in which the string looses its tension is high enough, on the order of $9.1D$ as estimated, then, around this value of $T$, the confining potential $V_{s}$ must be very strong, possibly preventing the freedom for the poles. With all these expectations, it would be important to investigate how the elementary excitations behave as a function of temperature. Our calculations is a first step in this direction.

\section{Results}

Now we present the results from Monte Carlo Simulations. The
calculations shown here are for lattices with sizes $10, 20,30,40,50,60$ and $70$ lattice spacings but in all figures we present only the results for lattice sizes $40,60, 70$. We start by presenting the results for the specific heat (see Fig.\ref{Heat}). We notice that, for all lattice sizes studied, the specific heat exhibits a sharp feature at a temperature $T_{p}$ approximately equal to $7.2D/k_{B}$. Indeed, the position of this peak does not seem to move as the lattice size $L$ is varied. On the other hand, its amplitude $C_{max}$ increases much slowly as $L$ increases. In the inset of Fig.\ref{Heat}, we show how $C_{max}$ behaves with $L$. Therefore, with the obtained data we expect a logarithmic  divergence of the specific heat in the thermodynamic limit. We also analyzed the pair density and the average separation between monopoles with opposite charges as a function of $T$. It is useful here to distinguish two types of monopoles: the less energetic ones in which the spins (in a vertex) are in the $3-in$, $1-out$ or $3-out$, $1-in$ states (here referred to as unit-charged monopoles) and the most energetic ones in which the spins are in the $4-in$ or $4-out$ states (doubly charged monopoles). Figure \ref{Density} shows the density of pairs containing monopoles with unitary charge ($\rho_{S}$) and also the density of pairs containing doubly charged monopoles ($\rho_{D}$, see the inset). They are calculated as the one-half of the thermodynamic average of the absolute value of the charge ($\pm 1$) and ($\pm 2$) respectively, summed over the lattice. For both cases, the density increases monotonously up to a maximum value achieved in the high-temperature limit.

\begin{figure}
\includegraphics[angle=0.0,width=8cm]{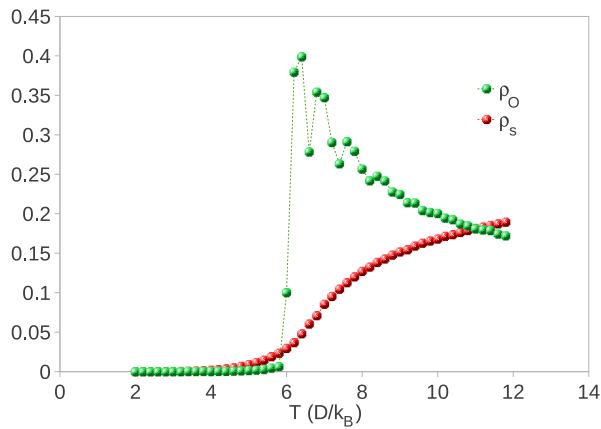}
\caption{ \label{Densityloop} (Color online) The density of string loops $4O$ ($\rho_{O}$) also exhibits a maximum around the temperature $T_{p} \simeq 7.2D$ (green balls). This defect carries no charge and is the second excited state. Just for comparison, the density of pairs with opposite charges ($\rho_{s}$) is also shown (red balls).}
\end{figure}

The size of the monopole pairs constitutes an internal degree of freedom, since the energy of a pair depends on
the distance between the members of the pair. Here we would like to know the average distance $r_{M}$ between two opposite poles as a function of temperature. Such a thermodynamic quantity may contain information about the possibility of monopoles separation and how they are organized into the system. For this calculation we consider only defects with unitary charges. The grouping of monopoles into pairs is unique as long as the distances between them are smaller than the average distance between the monopoles $r_{M}=1/\sqrt{\rho_{S}}$. As the size of the monopole pairs becomes larger than $r_{M}$, one would simply have to redefine the monopole pairs. The average size $r_{M}$ of the monopole pairs is calculated by using the method of assignment problems; it deals with the question of how to assign $n$ items (jobs, students) to $n$ other items (machines, tasks) \cite{Bukard}. In our case, we would like to assign $n$ positive charges to $n$ negative charges for a given configuration in such a way that the sum of distances of all possible pairing be a minimum. The results are shown in Fig.\ref{Separation}. The average separation has a local maximum at the same temperature $T_{p}$ in which the specific heat exhibits a peak ($ \sim 7.2 D/k_{B}$). We notice that the amplitude of this maximum increases slowly as the system size increases. Indeed, like the specific heat peak, the maximum in the average separation $d_{max}$ also increases  logarithmically with the system size $L$ ($d_{max} \propto \ln L$, see Fig.\ref{Maximum}) and hence, one could expect that a certain quantity of monopoles may be almost isolated for very large arrays. Indeed, in our simulations for temperatures $T \geq T_{p}$ considering lattices with $L\leq 80a$, we could observe some charges relatively distant from their respective counterparts (separated by distances of the order of $5a$). For instance, we show in Fig. \ref{Temp1} a distribution of positive (red circles) and negative (black circles) monopoles in a small lattice with $L=10a$ observed in our simulations for a temperature $T=6.0 D/k_{B}$ (i.e., below $T_{p}$). Note that there are few excitations and all monopoles with opposite charges are coupled by a string, forming pairs. On the other hand, Fig. \ref{Temp2} shows the same system for a temperature above $T_{p}$ ($T=7.6 D/k_{B}$). In this case, we see that a small quantity of monopoles are not connected by strings. In principle, they are free although some of them are not completely isolated (i.e., far away from other opposite poles). Furthermore, we also notice that some strings seem to be detached, not terminating in monopoles; there are few pieces of strings dispersed along the system (as said before, strings could be seen as lines which pass by adjacent vertices that obey the ice rule but sustaining topology $2$ rather than topology $1$). Of course, these figures exhibit only samples from a large number of data, but most of the data should be similar to the features of Fig.\ref{Temp1} for the regime of low temperatures and the features of Fig.\ref{Temp2} for the regime of high temperatures. Things must be clearer in the thermodynamic limit; in this case, some monopoles should become infinitely separated from their counterpart for temperatures $T \geq 7.2 D/k_{B}$. However, as the temperature is increased from zero, the monopole pair density grows simultaneously with an increase of the pair size (see also Fig.\ \ref{Density}). As the pairs become denser, there is less space to put in new pairs and hence the average pair size $r_{M}$ decreases for high temperatures. Really, we observe that, for $T < T_{p}$ the average separation $r_{M}$ does not depends on the lattice size $L$, while for $T \geq T_{p}$, this quantity has a tiny dependence on $L$ (at least in the range $7.2 D/k_{B}<T<12 D/k_{B}$). In this case, it is possible that monopoles may become completely isolated even for high temperatures ($T >T_{p}$) when $L\rightarrow \infty$. This picture for infinite systems corroborates the theoretical expectations for the existence of a phase with free monopoles \cite{Mol09} in large $2d$ artificial square ices, but the transition temperature ($\sim 7.2D/k_{B}$) should be little smaller than the estimated value $\sim 9.1D/k_{B}$ discussed earlier (remembering that the arguments of energy-entropy, in general, overestimate the correct quantity).

\begin{figure}
\includegraphics[angle=0.0,width=9cm]{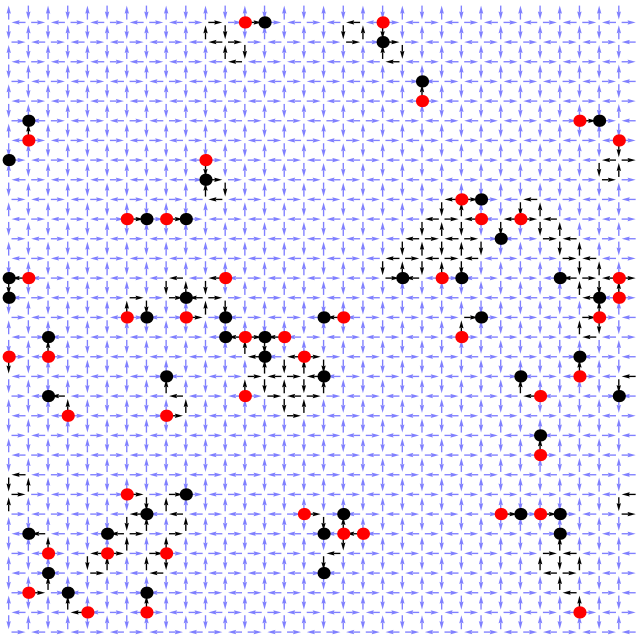}
\caption{ \label{LoopTseis} (Color online) A typical configuration of string loops of the type $4O$ for a temperature below $T_{p}$ (here, $T=6D/k_{B}$). At $T_{p}$, the number of $4O$ excitations proliferate in such way that a percolated cluster seems to be formed. The figure also shows the pairs of monopoles.}
\end{figure}

\begin{figure}
\includegraphics[angle=0.0,width=9cm]{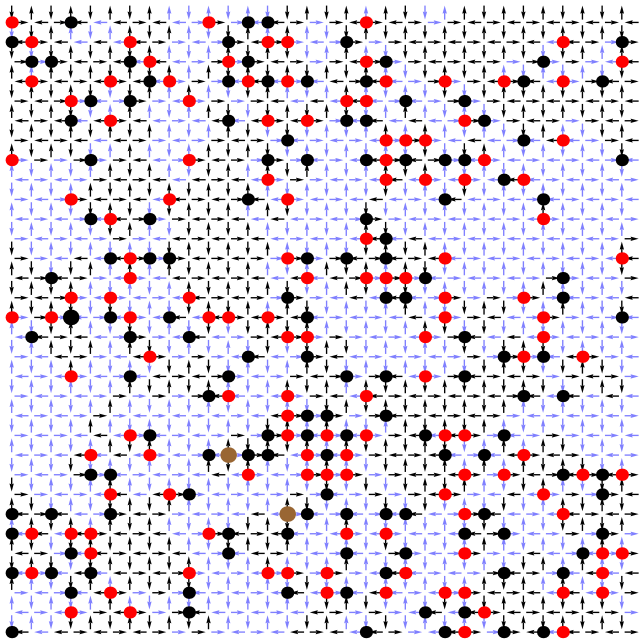}
\caption{ \label{LoopToito} (Color online) A typical configuration of string loops of the type $4O$ for a temperature above $T_{p}$ (here, $T=8D/k_{B}$). The figure also shows the pairs of monopoles.}
\end{figure}

We have also calculated the density of string loops $4O$, which is the defect with no charge but having the second lower energy (second excited state). Like the specific heat and the average separation, the density of defects $4O$ also displays a feature at $T_{p}$ (see Fig. \ \ref{Densityloop}). Note that the string loops $4O$ almost do not appear in the system for temperatures smaller than $T_{p}$. Indeed, they surge suddenly at $T_{p}$ and then, for temperatures above $T_{p}$, their number starts to decrease while the density of monopole pairs starts to increase more appreciable. Figures (\ \ref{LoopTseis}) and (\ \ref{LoopToito}) show typical distributions of defects $4O$ in the system for temperatures below and above $T_{p}$, respectively.

\section{Discussion}

In summary, assuming the spin-spin interaction to be purely dipole-dipole, we notice that, at a temperature $T_{p}$, there is a maximum in the mean separation of opposite monopoles that increases logarithmically with the system size $L$ ($d_{max} \propto \ln L$). Hence, the distance between  monopoles with opposite charges in the thermodynamic limit ($L \to \infty$) should diverge weakly, suggesting a possible unbinding of monopole pairs ($T<T_{p}$) into ''free" monopoles ($T>T_{p}$). However, to the authors knowledge, for a finite monopole density there is no diagnostic for (de)confinement based on a pair distribution function, for reasons analogous to the failure of the Wilson loop (which only knows perimeter laws in the presence of dynamical matter) to diagnose deconfinement in gauge theories. Indeed, from the three approaches that have been used to measure the static potential associated with the breaking of long flux tube between two quarks in QCD (i.e., correlation of Polyakov loops, variational ansatz and Wilson loops), string breaking has been seen only using the first two methods. On the other hand, the divergence found in $r_{M}$ could be understood in two different ways. It may be associated with either a vanishing string tension (which would lead to effectively free poles) or simply by the fact that in an order-disorder transition the correlation length (which is the only characteristic length of the system) diverges at the critical temperature. In this case, since the mean distance should be given in terms of the correlation length, then, it should also diverge. Of course, these two distinct ways to describe the system are closely related. We are faced thus with the question of the existence or not of a phase transition in this system.  If there is a phase transition, other question arises: what is its nature? It is worthy to note at this point \cite{Mol11} that although this system is closely related to the 16-vertex model, for which an exact solution is known, the range and symmetry of the interactions differ and thus we do not expect to observe the same critical behavior. Nevertheless, one point that deserves remark is the possible similarities between this system and the Ising model. In the two degenerated ground states, the $\sigma$ variables, related to the vorticity of each plaquette, can be seem as the spins of an antiferromagnetic (AF) Ising model. In the AF Ising model, as the temperature raises, clusters of flipped spins are found in the system and at the critical temperature one can find percolated clusters of spins. If there is some similarities between these systems one may expect thus that the $4O$ excitations, which can be viewed as flipped $\sigma$ variables, form clusters at low temperature that percolates at the critical temperature, justifying thus the increasing number of these excitations at the transition temperature. This picture is corroborated by the logarithmic divergence of the specific heat. Unfortunately, our results are not conclusive about the possibility of a phase transition, and much more work has to be done in order to answer this question. To try to put some extra light on the topic, we have also done some calculations restricting the islands interaction to nearest neighbors converging in the same vertex, which would lead to a kind of generalized $2d$ Ising system with the same ground state. Nevertheless, we have obtained that the vertices with topology $3$, in the $3$-in/$1$-out and $3$-out/$1$-in states, remain connected by strings (but now, there is no Coulomb interaction anymore). The interaction energy between two opposite vertices in topology $3$ (type $III$ vertices) is given by $b_{I}X +c_{I}$, where $b_{I}=26D/a$ and $c_{I}=34D$, much bigger than the usual results obtained for the long ranged dipolar interaction. Since the string tension remains, the arguments associated with the string configurational entropy should maintain valid and we have again the same problem as before (but with different energetics; for instance, the value of the temperature in which the quantities show a maximum changes to $16D/k_{B}$). Indeed, the specific heat, the average separation between opposite type $III$ vertices etc, have the same behavior found for the system with long-range dipolar interaction (not shown here).

 From a practical point of view, the divergence in $r_{M}$ in the thermodynamic limit, and thus the phase of large separation among monopoles should not be expected in finite systems. Due to the slow logarithmic divergence, the extrapolation of our results to a $2d$ lattice containing the Avogadro's number ($N_{a}^{2/3}=10^{16}=10^{8}\times 10^{8}$) of islands will imply in $d_{max} \sim 2.5a$ only. On the other hand, even with small values for $d_{max}$, some monopoles may become isolated for temperatures near $7.2D$ (see Fig. \ref{Temp2}). The challenge of building arrays using new materials (with an ordering temperature near room temperature ) and/or with reduced island volume and moment (and possibly with larger $L$) should be then an important issue for technological applications. Indeed, it concerns with the excitations evolution in these artificial compounds. These developments may experimentally determine the possibility of monopole dynamics, their lifetimes and so on. For instance, based only on the average separation results, we speculate that, near the temperature $T_{p}$,  the annihilation process of monopoles (without strings) should be more probable to occur in small arrays than in large arrays due to the fact that the mean separation between such opposite charges increases with the system size.

\ack{The authors thank CNPq, FAPEMIG, CAPES and FUNARBE (Brazilian agencies) for
financial support. We would like to thank Professors R. Moessner and G.M. Wysin for a careful reading of the manuscript and for helpful comments.}

\section*{References}


\end{document}